\documentclass[conference,a4paper]{IEEEtran}

\ifCLASSINFOpdf
\else
\fi
\usepackage[left=1.57cm,right=1.57cm,top=0.95cm,bottom=2.54cm]{geometry}
\usepackage{listings}
\usepackage{setspace}
\usepackage{graphicx}
\usepackage{tabularx,booktabs}
\usepackage{dingbat}
\usepackage{diagbox}
\usepackage{multirow} 
\usepackage[hyphens]{url}
\usepackage[hidelinks,breaklinks]{hyperref}
\usepackage{xcolor}
\usepackage{amsfonts, amsmath, amsthm, amssymb}
\usepackage{subcaption}
\hypersetup{breaklinks=true}
\usepackage{tikz}
\usepackage{textcomp}
\usepackage{lipsum}
\usepackage{svg}
\IEEEoverridecommandlockouts
\newcommand\copyrighttext{%
  \footnotesize \textcopyright 2026 IEEE.  Personal use of this material is permitted.  Permission from IEEE must be obtained for all other uses, in any current or future media, including reprinting/republishing this material for advertising or promotional purposes, creating new collective works, for resale or redistribution to servers or lists, or reuse of any copyrighted component of this work in other works.}
\newcommand\copyrightnotice{%
\begin{tikzpicture}[remember picture,overlay]
\node[anchor=south,yshift=10pt] at (current page.south) {\fbox{\parbox{\dimexpr\textwidth-\fboxsep-\fboxrule\relax}{\copyrighttext}}};
\end{tikzpicture}%
}
\usepackage{graphicx}
\graphicspath{{figures/}}
\begin{document}
\bstctlcite{IEEEexample:BSTcontrol}
%
\title{Sustainable Real-Time 8K60 HEVC Encoding for V2X: Repurposing Legacy NVENC Hardware at the Vehicular Edge}

\author{
    \IEEEauthorblockN{Kasidis Arunruangsirilert, Jiro Katto}
    \IEEEauthorblockA{Department of Computer Science and Communications Engineering, Waseda University, Tokyo, Japan
    \\\{kasidis, katto\}@katto.comm.waseda.ac.jp}
}
%

\maketitle

\copyrightnotice
\setstretch{0.90}
\begin{abstract}

The rapid advancement of Vehicle-to-Everything (V2X) communications and Tele-Operated Driving (ToD) demands ultra-low-latency, 8K60 video telemetry. However, deploying modern hardware at the vehicular edge is frequently hindered by supply chain constraints, high power budgets, and growing e-waste concerns.  This paper investigates a highly sustainable alternative: repurposing legacy NVIDIA Pascal GPUs for real-time 8K HEVC edge encoding. We demonstrate that triggering 2-Way Split Frame Encoding (SFE) on dual-NVENC GP104 and GP102 silicon successfully unlocks real-time 8K60 throughput with a negligible Rate-Distortion penalty of under 1\%. Crucially, our micro-architectural analysis reveals that smaller GPU dies significantly outperform larger flagship models in both raw throughput and energy efficiency. Because fixed-function encoding forces general-purpose Streaming Multiprocessor (SM) cores to sustain maximum frequencies while remaining idle, GPUs with fewer CUDA cores waste drastically less power. While benchmarking against the state-of-the-art RTX PRO 6000 Blackwell highlights a generational compression efficiency gap, Pascal’s functional HEVC architecture and native lack of B-frames align perfectly with ultra-low-latency V2X pipelines. Ultimately, repurposed mid-range Pascal GPUs present a highly capable, cost-effective, and e-waste mitigating solution for modern Intelligent Transportation Systems.

\end{abstract}



\setstretch{0.897}

%
\IEEEpeerreviewmaketitle

\vspace{-2mm}
\section{Introduction}

The rapid advancement of Vehicle-to-Everything (V2X) communications and autonomous teleoperation, often referred to as Tele-Operated Driving (ToD) \cite{10910110}, has introduced unprecedented data ingestion demands at the vehicular edge. High-fidelity remote driving, real-time monitoring, and edge-based computer vision increasingly rely on Ultra-High-Definition (UHD) video telemetry to maintain spatial awareness and minimize motion blur \cite{10217903}. Indeed, video transmission is one of the most bandwidth-consuming elements of Intelligent Transportation Systems (ITS), making efficient data compression strictly mandatory \cite{10574825}. With 8K resolution at 60 frames per second (8K60) emerging as a target standard, V2X applications inherently face the challenge of transmitting these massive data streams over wireless links. Because cellular uplink throughput, even in 5G Advanced (5G-A) deployments, remains a primary bottleneck that is highly subject to signal fluctuation \cite{11396852, 10467186}, deploying high-efficiency video compression at the edge is strictly necessary to prevent packet loss and maintain stable teleoperation \cite{8889542}.

Historically, deploying high-throughput edge nodes capable of real-time UHD compression presents significant energy and hardware challenges. Relying on software-based CPU encoding for UHD video introduces prohibitive latency and excessive power consumption. The transition to high-efficiency standards like High Efficiency Video Coding (HEVC) exponentially increases computational complexity over older codecs, making real-time CPU encoding too power-heavy for battery-constrained environments \cite{7763234,8081463}. This directly degrades electric vehicle (EV) battery efficiency and consumes critical compute headroom needed for other on-board autonomous tasks \cite{10637525,10910110}. Consequently, Graphic Processing Units (GPUs) with dedicated hardware video encoders, such as NVIDIA's Ada Lovelace or Blackwell GPUs, are typically mandated. However, as of early 2026, their widespread deployment at the edge is currently hindered by severe global supply chain constraints, widespread DRAM shortages, and prohibitive per-unit costs. Furthermore, the aggressive decommissioning of older hardware to make room for new accelerators exacerbates the growing environmental crisis of electronic waste (e-waste). As global e-waste generation surges, embracing sustainable computing practices, such as extending device lifespans and adopting circular economy models, has become essential to mitigate ecological damage \cite{10717432}. \looseness=-1

This paper investigates a highly sustainable, cost-effective alternative for V2X edge computing: re-purposing legacy NVIDIA Pascal GPUs with the sixth generation NVENC core, which supports full hardware-accelerated HEVC encoding. Specifically, we demonstrate that even low-power, single-NVENC-chip Pascal variants like the Quadro P600 (GP107) are highly capable edge encoders, successfully sustaining HEVC encoding at 8K resolution at 24 frames per second (8K24) for baseline telemetry, while consuming less than 20 W of power. For demanding 8K60 applications, we examine the GP104 and GP102 silicon, such as GTX 1070 and GTX 1080 Ti, respectively, which uniquely house dual-NVENC hardware encoders. While NVIDIA documentation formally associates advanced parallel encoding techniques with newer architectures, our research indicates that standard FFmpeg API calls can trigger Split Frame Encoding (SFE) on legacy dual-NVENC Pascal chips, unlocking real-time 8K60 HEVC encoding, prolonging its potential usability lifespan.

Furthermore, while the 6th-generation NVENC core natively lacks support for HEVC B-frames, its core HEVC implementation remains highly applicable for modern edge networks. Because ultra-low-latency V2X pipelines frequently disable B-frames to minimize bidirectional encoding and decoding delays, prioritizing immediate frame delivery over maximum compression efficiency \cite{10637525, 9253375}, the architectural limitations of the GTX-10 series do not preclude its use in modern, real-time vehicular implementations.

To validate the viability of re-purposing these legacy GPUs, this paper presents a comprehensive evaluation and comparative analysis. The contributions of this work are as follows:

\begin{itemize}
    \item We evaluate the encoding throughput, energy efficiency, and Rate-Distortion (RD) performance of the NVIDIA Pascal GPUs to determine their deployability in modern V2X scenarios.
    \item We quantify the compression efficiency trade-offs incurred when utilizing 2-Way SFE compared to a standard single encoder across the evaluated hardware.
    \item We benchmark the legacy Pascal GPUs against the state-of-the-art RTX Pro 6000 Blackwell GPU. By evaluating both High Quality (HQ) and Ultra High Quality (UHQ) modes, where NVIDIA claims Blackwell boasts a 5–10\% RD performance uplift over Ada Lovelace \cite{nvidia_2024b}, we compare the trade-off between two micro-architectures released nine years apart.
\end{itemize}

\section{Experiment Setup}

\subsection{Encoding Throughput Benchmark} \label{sec:thpt}

To isolate the peak encoding throughput of NVENC hardware and eliminate potential system-level bottlenecks (e.g., PCIe bandwidth, disk I/O latency, or system memory throughput), we utilized a synthetic frame generation methodology similar to previous works \cite{11396901, 11417632}. Using FFmpeg's \textit{lavfi} source filter, a black frame was generated and transferred to the GPU VRAM via the \textit{hwupload} filter. This single black frame was then looped internally within the GPU for the duration of the encoding session, ensuring that the measured frame rates represent the absolute upper bound of the NVENC hardware's computational capacity.

We evaluated the encoding throughput using the HEVC (H.265) 8-bit 4:2:0 format. To accurately reflect generational hardware limitations and capabilities, the encoder configurations were strictly delineated by architecture. For Pascal architecture, the 6th-generation NVENC lacks native support for HEVC B-frames, which prohibited the use of UHQ tuning mode; all Pascal tests were conducted with B-frames disabled and HQ tuning mode only. On the other hand, for Blackwell architecture, the frame structure was configured with 2 B-frames and 1 reference frame, which is typically acceptable in real-time encoding scenarios. We evaluated both the standard HQ and the newly introduced UHQ tuning modes. For both architectures, we tested across three distinct performance presets: P1 (Fastest), P4 (Medium), and P7 (Slowest/Highest Quality). In each trial, we encoded 14400 frames at 4K or 3600 frames at 8K resolution.

\subsection{Rate-Distortion Analysis}

\subsubsection{Test Datasets}

To evaluate compression efficiency across a diverse range of content characteristics, spatial complexities, and motion dynamics, we utilized three datasets:

\begin{enumerate}
  \item \textbf{Netflix Chimera  \cite{netflix}:} A DCI 4K (4096×2160) sequence at 59.94 fps consisting of 23 distinct scenes with a total length of 30:49 minutes.
  \item \textbf{ITE Ultra-High Definition Standard Test Sequences (Series A) \cite{ITE_2016}:} As described in the official documentation, this dataset includes 10 sequences at 4K and 11 sequences at 8K. All sequences feature a frame rate of 59.94 fps and are 15 seconds in length.
  \item \textbf{Twitch \cite{xiph.org}:} A 1080p (1920×1080) sequence at 60 fps consisting of 10 footages sourced from the most popular video games streamed on the Twitch platform, each with a runtime of 60 seconds.
\end{enumerate}

\subsubsection{Mezzanine File Preparation}

To facilitate efficient large-scale testing and allow for hardware-accelerated decoding (NVDEC) during the encoding trials, all source content was pre-processed into a high-fidelity mezzanine format. For the \textit{Netflix Chimera} dataset, which was originally provided in HDR, we performed a high-fidelity conversion to Standard Dynamic Range (SDR) and cropped the frame to a standard UHD resolution of 3840×2160. This was achieved via an FFmpeg filter chain that linearized the transfer characteristics, applied Hable tone-mapping to accurately preserve highlight details, and converted the color primaries to BT.709. As for the \textit{Twitch} dataset, the native 1080p sequences were upscaled to 4K UHD (3840×2160) with a Lanczos resampling filter to maintain fine details. Finally, for the \textit{ITE Ultra-High Definition} dataset, which is natively mastered in the BT.2020 wide color gamut, we applied a precise color space conversion to BT.709. This transformation explicitly preserved the limited (TV) video range during scaling to prevent luma and chroma clipping. Following these content-specific transformations, all sequences were encoded into a visually lossless mezzanine format using the software \textit{libx265} encoder with a Constant Rate Factor (CRF) of 10, standard 8-bit 4:2:0 planar YUV pixel format, and the \textit{medium} preset.

\subsubsection{Encoding Settings}

\begin{table}[!tbp]
\vspace{-2.5mm}
\setstretch{0.75}
\caption{Hardware and Software Configurations}
\centering
\label{tab:hardware}
\resizebox{8.5cm}{!}{\begin{tabular}{@{}ll@{}}
\toprule
\multicolumn{2}{c}{Encoding Systems}\\
\midrule
Hardware                 & Description  \\\midrule
Pascal GPUs& NVIDIA Quadro P600 (GP107, 
132 mm²)\\
& NVIDIA GeForce GTX 1060 3 GB (GP106, 
200 mm²)\\
& NVIDIA GeForce GTX 1070 (GP104, 
314 mm²)\\
& NVIDIA GeForce GTX 1080 Ti (GP102, 
471 mm²)\\
Blackwell GPU& NVIDIA RTX PRO 6000 Blackwell Max-Q \\
&Workstation Edition (GB202, 750 mm²)\\\midrule
Software & Version \\\midrule
OS & Microsoft Windows 11 Pro/Server 2025 Datacenter\\
ffmpeg & N-122268-g0dfaed77a6\\
OBS Studio & 32.0.4\\
GeForce Driver & GeForce Game Ready Driver 581.80 (581.80)\\
Quadro/RTX PRO Driver& NVIDIA RTX Driver Release 580 R580 U5 (581.80)\\\midrule
\multicolumn{2}{c}{VMAF/PSNR Calculation System}\\
\midrule
Hardware                 & Description  \\\midrule
CPU & Intel(R) Core(TM) Ultra 9 285K \\
RAM & Dual-Channel DDR5 128 GB (4×32 GB) @ 4400 MT/s \\ 
GPU & NVIDIA RTX PRO 5000 48GB Blackwell\\\midrule
Software & Version \\\midrule
OS & Ubuntu 24.04.3 LTS\\
ffmpeg & N-122271-g0629780cf6\\
libvmaf & v3.0.0 (b9ac69e6)\\
VMAF Model & vmaf\_4k\_v0.6.1neg \\
NVIDIA GPU Driver & 580.126.09 \\
NVIDIA CUDA Compiler & cuda\_13.1.r13.1\/compiler.36836380\_0 \\
\bottomrule
\end{tabular}}
\vspace{-2mm}
\end{table}

\begin{table}[!tbp]
\setstretch{0.85}
\caption{Target Encoding Bitrates for each resolution}
\vspace{-1.5mm}
\centering
\label{tab:bitrateRange}
\resizebox{7.5cm}{!}{\begin{tabular}{@{}lc@{}}
\toprule
Resolution                 & Bitrates (Mbps) \\\midrule
3840×2160 (2160p/4K) & 1, 2, 3, 4, 7, 10, 15, 22, 35, 50 \\
7680×4320 (4320p/8K) & 4, 6, 9, 14, 20, 30, 45, 70, 100, 150 \\
\bottomrule
\end{tabular}}
\vspace{-6.5mm}
\end{table}

We evaluated the codec/preset combinations defined in Section \ref{sec:thpt}, adhering to the specific generational hardware limitations (e.g., the lack of B-frames support on Pascal). The specific hardware and software configurations for the systems used in this study are detailed in Table \ref{tab:hardware}. For the encoding systems, we utilized multiple nodes with different GPU models to avoid swapping GPUs and system reassembly between runs. We closely follow the real-time broadcast and streaming requirements to ensure that our experiment represents the actual use cases. The following encoding parameters were applied across all runs:

\begin{itemize}
    \item \textbf{Rate Control:} Constant Bitrate (CBR) with a Video Buffer Verifier (VBV) buffer set to 2× the target bitrate.
    \item \textbf{GOP Structure:} Fixed 2-second closed Group of Pictures (GOP) without scene cut detection.
    \item \textbf{Format:} HEVC (H.265) Main profile, 4:2:0 chroma subsampling, 8-bit depth.
    \item \textbf{Optimization:} Spatial and Temporal Adaptive Quantization were disabled, and lookahead was disabled to strictly minimize latency.
\end{itemize}

Crucially, while the RTX PRO 6000 Blackwell supports up to 4-Way SFE, we restrict our SFE evaluation strictly to 1-Way (No SFE) and 2-Way SFE. Deploying a flagship GPU with a tri- or quad-NVENC chip (such as the RTX PRO 6000 Blackwell) into automotive environments is economically, physically, and thermally prohibitive. Because practical, mid-range Commercial Off-The-Shelf (COTS) GPUs viable for vehicular edge nodes typically feature a maximum of two NVENC cores, evaluating 4-Way SFE falls entirely outside the scope of applied V2X deployments. To generate accurate Rate-Distortion curves, we encoded each sequence across ten bitrate points spanning from 1 Mbps to 50 Mbps for 4K content, and 4 Mbps to 150 Mbps for 8K content (see Table \ref{tab:bitrateRange}). \looseness=-2

\subsubsection{Quality Metric}

\begin{table}[!tbp]
\setstretch{0.7}

\caption{Encoding Throughput (FPS) by Configuration}
\vspace{-2mm}
\centering
\label{tab:EncodingThpt}
\resizebox{7cm}{!}{\begin{tabular}{@{}lcccccc@{}}
\toprule
\multirow{2.5}{*}{GPU Model}& \multicolumn{3}{c}{No SFE} & \multicolumn{3}{c}{2-Way SFE} \\
\cmidrule(lr){2-4} \cmidrule(lr){5-7}
& P1 & P4 & P7 & P1 & P4 & P7 \\
\midrule
\multicolumn{7}{c}{4K UHD (2160p)}\\\midrule
Quadro P600&106&94&93&\multicolumn{3}{c}{N/A}             \\
GTX 1060 3 GB&121&108&106&\multicolumn{3}{c}{N/A}   \\
GTX 1060 3 GB (OC)&136&121&118&\multicolumn{3}{c}{N/A}   \\
GTX 1070&123&110&108&227&204&201  \\
GTX 1070 (OC)&135&121&119&247&223&219  \\
GTX 1080 Ti&116&104&102&217&194&191  \\
GTX 1080 Ti (OC)&123&110&108&230&208&202  \\
RTX PRO 6000 (HQ)&247&125&39&451&232&76    \\
RTX PRO 6000 (UHQ)&110&82&34&111&79&37      \\
\midrule
\multicolumn{7}{c}{8K UHD (4320p)}\\\midrule
Quadro P600&27&24&23&\multicolumn{3}{c}{N/A}        \\
GTX 1060 3 GB&31&27&27&\multicolumn{3}{c}{N/A}   \\
GTX 1060 3 GB (OC)&34&31&30&\multicolumn{3}{c}{N/A}   \\
GTX 1070&32&28&28&62&55&54  \\
GTX 1070 (OC)&35&31&31&68&60&60  \\
GTX 1080 Ti&30&27&27&59&54&52  \\
GTX 1080 Ti (OC)&33&29&29&63&57&55  \\
RTX PRO 6000 (HQ)&66&32&10&125&62&20 \\
RTX PRO 6000 (UHQ)&48&26&10&55&32&15  \\
\bottomrule
\end{tabular}}
\vspace{-2mm}
\end{table}

\begin{table}[!tbp]

\setstretch{0.75}
\vspace{-0.5mm}
\caption{Average Core Clock (MHz) (8K HQ P7)}
\vspace{-2mm}
\centering
\label{tab:clockSpeed}
\resizebox{5.5cm}{!}{\begin{tabular}{@{}lcc@{}}
\toprule
GPU Model& No SFE & 2-Way SFE \\
\midrule
Quadro P600 & 1620 & N/A \\
GTX 1060 3 GB&1847.5& N/A \\
GTX 1060 3 GB (OC)&2062.5& N/A\\
GTX 1070&1923.5&1923.5\\
GTX 1070 (OC)&2100.5&2100.5\\
GTX 1080 Ti&1961.5&1961.5\\
GTX 1080 Ti (OC)&2100.5&2100.5\\
RTX PRO 6000 (HQ)&2475&2475\\
RTX PRO 6000 (UHQ)&2475&2475\\
\bottomrule
\end{tabular}}
\vspace{-6mm}
\end{table}

Rate-distortion performance was evaluated using the Bjøntegaard Delta Rate (BD-Rate) metric, derived from the ten distinct bitrate points to ensure curve accuracy (see Table \ref{tab:bitrateRange}). We evaluated objective video quality using two primary metrics: the standard Peak Signal-to-Noise Ratio calculated on the luminance channel (PSNR-Y), which is the standard metric commonly found in video encoding literature, and the Video Multimethod Assessment Fusion 4K (VMAF 4K) score. To expedite the massive computational workload associated with multi-resolution objective assessment, VMAF was calculated using the hardware-accelerated \textit{libvmaf\_cuda} implementation on a dedicated quality calculation system, the specifications of which are detailed alongside the encoding hardware in Table \ref{tab:hardware}. For all VMAF evaluations, we utilized the \textit{vmaf\_4k\_v0.6.1neg} model, as it is specifically optimized to isolate pure compression artifacts and penalize artificial enhancement.

\subsection{Power Consumption and Overclocking}

To assess the power consumption characteristics of the GPUs under sustained encoding loads, we monitored the Board Power Draw and GPU Chip Power Draw. Telemetry data was logged using GPU-Z. It should be noted that the Quadro P600 does not report isolated GPU chip power consumption via standard telemetry APIs; therefore, only its total board power draw was recorded.

Additionally, official NVIDIA documentation states that NVENC performance varies across GPU classes but ``scales (almost) linearly with the clock speeds for each hardware" \cite{nvidia_corporation_2026}. To empirically evaluate this claim and maximize the encoding throughput of the legacy Pascal architecture for edge deployments, we utilized MSI Afterburner to aggressively overclock the consumer-grade Pascal GPUs. We pushed the core clocks of both the GTX 1080 Ti and the GTX 1070 to a sustained 2100 MHz during the overclocked encoding experiments. Because the Quadro P600 does not support core overclocking, it was evaluated only at its stock frequency of 1620 MHz. For each configuration, the encoding workload was sustained for a fixed duration of 30 seconds. This duration was selected to negate initial boost-clock thermal transients and ensure the GPU had reached a representative thermal steady state. Telemetry was logged at 1-second intervals throughout the test, from which the average power draw for each parameter was calculated and reported.

\section{Results and Analysis}

\subsection{Encoding Throughput} \label{sec:EncThpt}

\begin{table}[!tbp]
\setstretch{0.8}
\caption{BD-Rate: Enabling Split Frame Encoding (\%)}
\vspace{-2mm}
\centering
\label{tab:SFEPenalty}
\resizebox{8.7cm}{!}{\begin{tabular}{@{}lccccccccc@{}}
\toprule
\multirow{2.5}{*}{Dataset}& \multicolumn{3}{c}{Pascal} & \multicolumn{3}{c}{Blackwell HQ} & \multicolumn{3}{c}{Blackwell UHQ}\\
\cmidrule(lr){2-4} \cmidrule(lr){5-7} \cmidrule(lr){8-10}
& P1 & P4 & P7 & P1 & P4 & P7 & P1 & P4 & P7 \\
\midrule

\multicolumn{10}{c}{BD-Rate PSNR-Y}\\\midrule
Netflix Chimera&1.07&1.05&0.84&4.06&2.91&2.57&2.84&3.34&3.43\\
ITE 4K&1.02&0.96&1.03&0.79&0.98&0.79&0.97&1.10&1.23\\
ITE 8K&0.30&0.34&0.26&0.71&0.99&0.90&0.13&0.73&0.85\\
Twitch&1.49&1.68&1.34&2.84&3.51&3.52&3.88&3.76&3.72\\
\midrule
\textbf{Average}&0.97&1.01&0.87&2.10&2.10&1.94&1.95&2.23&2.31\\\midrule
\multicolumn{10}{c}{BD-Rate VMAF 4K}\\\midrule
Netflix Chimera&0.91&0.92&0.99&4.88&7.33&4.35&4.03&4.88&5.02\\
ITE 4K&0.53&0.52&0.61&0.44&0.65&0.43&0.82&1.14&1.06\\
ITE 8K&-0.32&-0.12&-0.06&0.49&0.53&0.47&-0.23&0.59&0.72\\
Twitch&1.41&1.64&1.20&3.15&3.87&3.95&4.56&3.96&4.05\\
\midrule
\textbf{Average}&0.63&0.74&0.68&2.24&3.09&2.30&2.29&2.64&2.71\\

\bottomrule
\end{tabular}}
\vspace{-6mm}
\end{table}

\begin{table*}[!tbp]
\setstretch{0.75}
\caption{BD-Rate: Pascal vs Blackwell Rate-Distortion Performance Across Various Configurations (\%)}
\vspace{-2mm}
\centering
\label{tab:BDRate}
\resizebox{13cm}{!}{\begin{tabular}{@{}lcccccccccccc@{}}
\toprule
\multirow{2.5}{*}{Dataset}& \multicolumn{3}{c}{Blackwell HQ} & \multicolumn{3}{c}{Blackwell UHQ} & \multicolumn{3}{c}{Blackwell HQ (SFE)} & \multicolumn{3}{c}{Blackwell UHQ (SFE)}\\
\cmidrule(lr){2-4} \cmidrule(lr){5-7} \cmidrule(lr){8-10} \cmidrule(lr){11-13}
& P1 & P4 & P7 & P1 & P4 & P7 & P1 & P4 & P7 & P1 & P4 & P7\\
\midrule
\multicolumn{13}{c}{BD-Rate PSNR-Y}\\\midrule
Netflix Chimera&-12.96&-10.96&-11.22&-24.46&-25.98&-25.96&-10.16&-9.02&-9.65&-22.86&-23.94&-23.79   \\
ITE 4K&-20.17&-21.65&-21.17&-31.47&-32.62&-32.50&-20.17&-21.51&-21.21&-31.38&-32.53&-32.46 \\
ITE 8K&-18.07&-19.49&-19.29&-32.32&-34.10&-34.13&-17.72&-18.88&-18.74&-32.43&-33.85&-33.85 \\
Twitch&-9.31&-8.46&-7.93&-18.41&-19.35&-19.62&-8.07&-6.72&-5.87&-16.35&-17.41&-17.42       \\
\midrule
\textbf{Average}&-15.13&-15.14&-14.90&-26.66&-28.01&-28.05&-14.03&-14.03&-13.87&-25.76&-26.93&-26.88 \\
\midrule
\multicolumn{13}{c}{BD-Rate VMAF 4K}\\\midrule
Netflix Chimera&-0.20&-0.20&2.20&-13.96&-16.02&-15.90&4.52&7.84&6.44&-10.69&-11.98&-11.78           \\
ITE 4K&-10.65&-12.79&-12.31&-17.88&-20.33&-20.29&-10.80&-12.68&-12.45&-17.66&-19.95&-20.16 \\
ITE 8K&-8.71&-10.15&-9.88&-20.41&-23.22&-23.21&-7.89&-9.53&-9.40&-20.32&-22.66&-22.76      \\
Twitch&0.47&-0.51&0.16&-7.15&-10.21&-10.69&2.21&1.78&2.98&-4.17&-7.81&-7.85                \\
\midrule
\textbf{Average}&-4.77&-5.91&-4.96&-14.85&-17.45&-17.53&-2.99&-3.15&-3.11&-13.21&-15.60&-15.64       \\
\bottomrule
\end{tabular}}
\vspace{-6.5mm}
\end{table*}

Table \ref{tab:EncodingThpt} and Table \ref{tab:clockSpeed} present encoding throughputs and sustained clock speeds, respectively. When evaluating baseline single-encoder performance at 8K resolution, the legacy Pascal architecture falls short of the 60 fps threshold. The single-NVENC Quadro P600 achieves a baseline of 27 fps at the P1 (Fastest) preset, while the GTX 1070 and GTX 1080 Ti hover between 30 and 32 fps. However, activating 2-Way SFE on the dual-NVENC GP104 (GTX 1070) and GP102 (GTX 1080 Ti) silicon yields a near-linear performance scaling of approximately 1.9×. Under 2-Way SFE, the GTX 1070 jumps to 62 fps at 8K P1, successfully crossing the 60 fps real-time threshold for UHD teleoperation.

Correlating the clock speeds in Table \ref{tab:clockSpeed} with the throughput data reveals a distinct micro-architectural efficiency trend tied to GPU die size. Although the GTX 1080 Ti (GP102) represents the flagship consumer silicon of the Pascal generation and operates at a higher stock clock (1961.5 MHz) than the GTX 1070 (1923.5 MHz), it consistently underperforms the smaller GP104 die in raw encoding throughput. This counterintuitive result can be attributed to the physical dimensions of the silicon: the larger GP102 die inherently necessitates longer internal electrical paths. For highly serialized, memory-intensive fixed-function tasks like video encoding, this increased physical distance introduces higher signal propagation latency, which bottlenecks the NVENC cores and reduces overall performance despite the frequency advantage. \looseness=-1

To evaluate NVIDIA's claim that NVENC performance scales linearly with clock speed, we overclocked both GPUs' core clock to a sustained 2100.5 MHz. Even when locked to the exact same core clock frequency, the smaller-die GTX 1070 achieved 68 fps (8K P1, 2-Way SFE) compared to the larger-die GTX 1080 Ti's 63 fps. This demonstrates that the smaller GP104 die yields a notably higher FPS-per-clock ratio for fixed-function encoding. Consequently, when repurposing legacy hardware for V2X edge nodes, sourcing the smaller-die GP104 GPUs equipped with dual NVENCs is not only more cost-effective and power-efficient but also actively preferable for maximizing encoding throughput.

When benchmarking these Pascal GPUs against the RTX PRO 6000 Blackwell, a distinct micro-architectural divergence emerges. Using HQ tuning, the Blackwell architecture demonstrates significantly higher throughput, delivering 125 fps at 8K P1 using 2-Way SFE. However, when Blackwell's new UHQ tuning mode is enabled, encoding throughput degrades severely. At 4K resolution (P1), Blackwell GPU with UHQ tuning achieves 110 fps and 111 fps with SFE disabled and enabled, respectively. At 8K resolution (P1), it scales poorly from 48 fps without SFE to only 55 fps with SFE enabled. As detailed in \cite{abhijit_patait_2024}, this profound lack of parallel scaling occurs because the UHQ preset alters the encoding pipeline into a hybrid mode between the dedicated NVENC silicon and the general-purpose Streaming Multiprocessor (SM) cores. By requiring frames to be processed by the SM cores for advanced rate-distortion optimizations prior to hardware encoding, the architecture introduces a severe system-level serialization bottleneck that starves the NVENC engines. \looseness=-1

Consequently, an overclocked, repurposed GTX 1070 leveraging 2-Way SFE, which is unofficially supported, actually outperforms a flagship Blackwell GPU operating in UHQ tuning mode for 8K encoding. This validates the dual-NVENC GP104 architecture not merely as a viable fallback, but as a highly capable and sustainable COTS solution capable of sustaining real-time 8K60 vehicular telemetry.

\subsection{Rate-Distortion (RD)}

\begin{table}[!tbp]
\setstretch{0.8}
\vspace{1.4mm}
\caption{BD-Rate: 2-Way SFE Pascal vs Blackwell HQ P1 without SFE (\%)}
\vspace{-2mm}
\centering
\label{tab:RDThpt}
\resizebox{7.5cm}{!}{\begin{tabular}{@{}lcccccc@{}}
\toprule
\multirow{2.5}{*}{Dataset}& \multicolumn{3}{c}{BD-Rate PSNR-Y} & \multicolumn{3}{c}{BD-Rate VMAF 4K} \\
\cmidrule(lr){2-4} \cmidrule(lr){5-7}
& P1 & P4 & P7 & P1 & P4 & P7 \\
\midrule
Netflix Chimera&-13.94&-8.75&-7.27&-1.10&3.34&4.57     \\
ITE 4K&-20.84&-16.99&-15.88&-11.16&-7.65&-6.48\\
ITE 8K&-18.32&-14.93&-13.81&-8.45&-5.39&-4.14 \\
Twitch&-10.64&-5.94&-4.16&-0.96&3.08&5.33     \\
\midrule
\textbf{Average}&-15.93&-11.65&-10.28&-5.42&-1.65&-0.18 \\
\bottomrule
\end{tabular}}
\vspace{-6mm}
\end{table}

\subsubsection{Generational Rate-Distortion Evolution: Pascal vs. Blackwell}

Table \ref{tab:BDRate} presents the BD-Rate comparison,  benchmarking all Blackwell configurations against the legacy Pascal architecture. First, in a single-encoder configuration, we compare corresponding presets between Pascal and Blackwell. Blackwell's HQ mode delivers a substantial improvement, achieving an average BD-Rate PSNR-Y saving of approximately 15.1\% across the P1, P4, and P7 presets over Pascal. This gain is even more pronounced in the hybrid UHQ mode, which achieves an average BD-Rate PSNR-Y saving of over 27.5\% against Pascal at the same presets. The BD-Rate VMAF 4K metric paints a similar picture, with Blackwell HQ providing a 5.2\% average saving and UHQ providing 16.6\% average saving. While the difference is significant, it should be noted that the performance gap is partially the result of Pascal GPUs lacking B-frame support, and a smaller gap should be expected in B-frame-prohibited use cases.

When both architectures are configured with 2-Way SFE, the generational advantage is largely maintained. When comparing SFE Pascal to SFE Blackwell, the latter continues to demonstrate superior RD performance. As shown in the Table \ref{tab:BDRate}, the SFE-enabled Blackwell Configuration provides an average BD-Rate PSNR-Y saving of 14.0\% relative to the Pascal baseline. Considering that enabling SFE on Pascal introduces a minor $\approx$1\% penalty (see Table \ref{tab:SFEPenalty}),  the effective generational gain of SFE Blackwell over SFE Pascal remains consistently high. On the other hand, the UHQ mode with SFE maintains its commanding lead, delivering an average PSNR-Y saving of 26.5\% against the Pascal architecture, also with SFE enabled.

\subsubsection{Rate-Distortion Penalty of Split-Frame Encoding}

While SFE is crucial for achieving high encoding throughput, splitting the frame into two slices prevents the encoder from exploiting vertical spatial redundancies across slice boundaries, necessarily incurring a compression efficiency penalty. Table \ref{tab:SFEPenalty} quantifies this BD-Rate overhead when moving from a single encoder to a 2-Way SFE configuration. For the Pascal architecture, this trade-off is exceptionally favorable. Enabling 2-Way SFE incurs a negligible average penalty of 0.95\% in BD-Rate PSNR-Y and 0.68\% in BD-Rate VMAF 4K across all presets. This demonstrates that for a minimal loss in quality, throughput can be nearly doubled, making SFE a highly efficient operational mode for Pascal.

In contrast, the penalty on the Blackwell architecture is much more severe. In HQ mode, enabling 2-Way SFE results in an average overhead of 2.05\% (PSNR-Y) and 2.54\% (VMAF 4K).  This suggests that Blackwell’s more advanced encoding pipeline is more sensitive to the loss of cross-slice redundancies. The penalty is similar for UHQ mode, averaging 2.16\% for PSNR-Y and 2.55\% for VMAF across all three tested presets. While still a modest penalty, it is roughly double that observed on Pascal, highlighting a clear architectural trade-off between generational efficiency gains and sensitivity to parallelization penalty.

\subsubsection{ Performance Analysis at Throughput Parity: 2-Way SFE Pascal vs. Blackwell HQ}

As established in Section \ref{sec:EncThpt}, a Pascal GPU using 2-Way SFE delivers encoding throughput comparable to that of a Blackwell GPU running in a single encoder HQ configuration. This allows for a direct, throughput-equivalent RD performance comparison, with the results detailed in Table \ref{tab:RDThpt}. At the P1 preset, the Blackwell HQ single-encoder configuration demonstrates superior RD performance. It achieves an average BD-Rate PSNR-Y saving of a remarkable 15.93\% and an average BD-Rate VMAF 4K saving of 5.42\% compared to the legacy Pascal GPU running with 2-Way SFE. This means that to achieve the same objective quality, the modern Blackwell single-encoder requires significantly less bitrate than the dual-encoder Pascal setup, even when both are outputting frames at a similar rate.

This efficiency gap narrows but remains substantial as encoding complexity increases. At the P4 preset, Blackwell’s BD-Rate PSNR-Y advantage remains a significant 11.65\%. By the P7 (Highest Quality) preset, the PSNR-Y advantage is still a considerable 10.28\%, although the VMAF 4K performance becomes statistically equivalent (-0.18\%). 

Therefore, while Pascal GPU with 2-Way SFE enabled can successfully match the encoding throughput of Blackwell's single encoder configuration, the modern Blackwell architecture is fundamentally more efficient. However, this RD performance gap narrows considerably at higher quality presets. Notably, stepping up to the P7 preset on Pascal completely eliminates the BD-Rate VMAF 4K penalty while incurring only a minor throughput penalty, a stark contrast to the severe performance degradation of high-quality presets on modern architectures. Hence, the P7 preset is a highly recommended configuration for repurposed Pascal deployments, offering a compelling balance that maximizes rate-distortion performance for a negligible cost in real-time throughput.

\vspace{-1mm}

\subsection{Power Consumption and Energy Efficiency}

\begin{table}[!tbp]
\setstretch{0.8}

\caption{Average Power Consumption (W) (8K HQ P7)}
\vspace{-2mm}
\centering
\label{tab:PowerConsumption}
\resizebox{7.5cm}{!}{\begin{tabular}{@{}lcccc@{}}
\toprule
\multirow{2.5}{*}{GPU Model}& \multicolumn{2}{c}{Board} & \multicolumn{2}{c}{Chip} \\
\cmidrule(lr){2-3} \cmidrule(lr){4-5}
& No SFE & 2-Way & No SFE & 2-Way \\
\midrule
Quadro P600&14.3&N/A&N/A&N/A             \\
GTX 1060 3 GB&42.3&N/A&24.2&N/A             \\
GTX 1060 3 GB (OC)&43.8&N/A&25.8&N/A             \\
GTX 1070&51.2&54.2&23.4&24.1 \\
GTX 1070 (OC)&52.7&57.1&24.3&26.8 \\
GTX 1080 Ti&88.3&90&37.6&38.6   \\
GTX 1080 Ti (OC)&90.7&93.5&39.9&41.8 \\
RTX PRO 6000 (HQ)&86.7&87&33.3&33.5   \\
RTX PRO 6000 (UHQ)&94.1&94.1&36.2&36.2 \\
\bottomrule
\end{tabular}}
\vspace{-6mm}
\end{table}

To assess operational viability in power-constrained vehicular edge environments, Table \ref{tab:PowerConsumption} shows the sustained power draw during an 8K P7 encoding workload. It was found that activating a second NVENC core via 2-Way SFE incurs a nearly negligible power penalty (e.g., GTX 1070 chip power increases by only 0.7 W). This occurs because the baseline power draw is dominated by the GPU exiting its low-power state and sustaining maximum clock speeds to drive the NVENC silicon. Consequently, the general-purpose SM cores are forced to run at maximum frequency, wasting power despite remaining idle during hardware-accelerated encoding. 

Because this high-performance SM state dominates power consumption, GPUs with fewer SM/CUDA cores are inherently more power-efficient for fixed-function encoding workloads. At the ultra-low-power end, the single-NVENC Quadro P600, equipped with only 384 CUDA cores, operates at an exceptionally lean 14.3 W total board power, making it ideal for sustainable 8K24 telemetry. For 2-Way SFE capable cards, the GTX 1070 (1920 CUDA cores) draws 54.2 W, whereas the flagship GTX 1080 Ti (3584 CUDA cores) demands 90 W to perform the exact same task. By wasting significantly less power on idle SM cores while simultaneously delivering higher raw encoding throughput, the GTX 1070 (GP104) drastically eclipses the flagship GP102 silicon in performance-per-watt. Even when aggressively overclocked to sustain 60 fps, the GTX 1070 (OC) board's power rises to only 57.1 W, cementing it as an optimally sustainable V2X solution. \looseness=-1

Finally, comparing these legacy GPUs to the modern RTX PRO 6000 Blackwell highlights the energetic cost of hybrid encoding modes on a fundamentally larger architecture. Despite its massive die size and 24064 CUDA cores, the Blackwell GPU draws 87 W of board power in standard HQ mode. However, shifting to the new UHQ mode elevates this to 94.1 W. This 7.1 W penalty supports our earlier analysis: UHQ forces this vast array of SM cores to actively wake up and perform rate-distortion optimizations. Because this SM dependency actively consumes more power while severely reducing throughput, UHQ is an unviable configuration for real-time, energy-conscious vehicular deployments. 

\section{Conclusion and Future Work}

In this paper, we investigated a highly sustainable and cost-effective approach for V2X edge computing by repurposing legacy NVIDIA Pascal GPUs for real-time 8K HEVC video telemetry. While NVIDIA claimed that SFE is not supported on Pascal GPUs, we demonstrated that the same NVIDIA Video Codec SDK API calls used in modern NVIDIA GPUs can trigger 2-Way SFE on legacy dual-NVENC GP104 and GP102 silicon, successfully unlocking real-time 8K60 HEVC encoding capabilities with a negligible RD penalty of less than 1\%. Our architectural analysis revealed a critical micro-architectural efficiency trend: smaller GPU dies with fewer SM cores inherently outperform larger dies in flagship models, both in terms of raw encoding throughput and energy efficiency. Because fixed-function hardware encoding forces the GPU out of its low-power state—running the SM cores at maximum frequency even while they remain idle—minimizing the total CUDA core count drastically reduces wasted power. Consequently, smaller GP104 dies offer a vastly superior performance-per-watt ratio and reduced signal propagation latency compared to larger GP102 silicon.

Although benchmarking against Blackwell GPU reveals a significant generational gap in RD performance, the legacy NVENC remains a fully functional and highly practical HEVC encoder. This compression efficiency difference is entirely workable within the parameters of vehicular telemetry. Furthermore, Pascal's native lack of HEVC B-frame support is not a practical disadvantage in this domain; ultra-low-latency V2X and teleoperation pipelines frequently disable B-frames by default to strictly minimize bidirectional encoding and decoding delays. Ultimately, repurposed mid-range Pascal GPUs configured with 2-Way SFE at the P7 preset present a highly capable, e-waste mitigating solution for modern Intelligent Transportation Systems.

For future work, we will investigate how different die sizes and clock speed curves contribute to NVENC encoding throughput and compare these micro-architectural dynamics to application-specific Video Processing Units (VPUs) for power-constrained edge environments.

\section*{Acknowledgement}

This paper is supported by the Ministry of Internal Affairs and Communications (MIC) Project for Efficient Frequency Utilization Toward Wireless IP Multicasting and the Japan Science and Technology Agency (JST) CRONOS Grant Number JPMJCS25N2.




%
\setstretch{0.80}
\Urlmuskip=0mu plus 1mu\relax
\bibliographystyle{IEEEtran}
\bibliography{b_reference}

\end{document}